\documentclass[a4paper,11pt]{article}
\usepackage[cp1251]{inputenc}
\usepackage[russian]{babel}
\usepackage[dvips]{graphicx}
\begin{document}
\Large
\begin{center}
\textbf{A.N. Sazonov\\
WBVR Observations of the Variable Star  V1341 Cyg in 1986-1992}\\

Sternberg Astronomical Institute, Universitetskii pr. 13, Moscow, 119992 Russia
\end{center}
\vspace{12pt}

\section*{Introduction}

~ We report the results of WBVR photometry of the massive close x-ray binary V1341 Cyg=Cyg X-2
in 1986-1992. In the "off" state the amplitudes of the periodic component of  light variations
lie in the  $0^{m}.265 - 0^{m}.278$ and  $0^{m}.290 - 0^{m}.320$ intervals for the B- and W-bands,
respectively, and is determined by the ellipsoidal shape of the optical component, which is distorted
by the gravitational influence of its x-ray companion. Some of the active states of the system
(long outbursts) may be due to the instability of the accretion disk (AD) and, possibly, to the
nonstationarity of the gaseous flows and other accretion formations. The close binary system (CBS)
contains a low-luminosity AD.
The light curves for the observing seasons considered exhibit no signs of eclipses in the vicinity of
the phases of the upper and lower conjunctions either in the quiescent or in the active state.
~ We are the first to calibrate the photometry of the comparison and check stars against the
WBVR catalog via  JHK magnitudes. The photometric reduction errors  with this calibration are
equal to  3\% for the  B and V bands and 8-10\% for the  W and\textit{} R\textit{} bands.
~ In this paper we use the refined estimates for the neutron-star mass, Keplerian-orbit radius,
and new estimates for the distance to Cyg X-2 computed in terms of the relativistic precession
model (RPM)~\cite{Smale1998}, which apply to different neutron-star states adopted from~\cite{Miller1998}.
We conclude, based on an analysis of our photometric data with the above corrections applied, that
the optical component in the close binary V1341 Cyg=Cyg X-2 is a red-giant star and not a  "blue
straggler", as some authors suggest.

~ This work is of current interest because we: (1) analyze long-period variability of the system over seven
years and (2) compare our optical observations with Ginga x-ray data obtained during
the observing periods considered.

\section*{Observations}

~ We computed the transformation coefficients during every observing season based on the
data for the best photometric nights. We corrected our broadband (heterochromatic) WBRI photometry of
Cyg X-2=V 1341 Cyg for atmospheric extinction.

~ The results of our optical observations of the CBS V 1341 Cyg = Cyg X-2 are listed in Table~4 (which is
available in electronic form at {http://lnfm1.sai.msu.ru/$\sim$sazonov/~Cyg~X-2}).

~ The observed 1986-1992 light curves of the system agree qualitatively with the optical
light curves of other observers. The system was in a quiescent state. Optical data partly correlate with x-ray
data for the same observing periods~\cite{Hasinger1989}.

\section*{Results of Observations and their Interpretation}

~ The optical light curves of the system in the active and quiescent states differ appreciably by, among other
things, extra flux contribution from the AD (when the system is in the active or intermediate
state).

~ Regular cyclic low-amplitude outbursts  (the first kind of activity), which last several
days, appear to be due to the interaction of gaseous flows emerging from the companion and to the instability
of accretion and that of the outer parts of the AD of the compact relativistic object of
the CBS.

~ The second kind of activity has the form of rapid outbursts lasting less than one day, which are naturally
attributed to the activity of the x-ray source. A typical example of such a rapid outburst of the second kind
was observed over 60 minutes on JD2446735.+, when the brightness increased by  0$^{m}$.04 in the W band,
by~0$^{m}$.03 in the B and V bands, and 0$^{m}$.02 in the R band. Deviations of the flux from the mean level
do not exceed $3\sigma$.

~ UV flux outbursts are observed in the orbital phase intervals $\varphi$=0$^p$.3 and $\varphi$=0$^p$.8 even
when the system is in the quiescent state.

\section*{Rapid Variability}

~ We studied the rapid variability of the star in the WBVR bands at minimum and maximum light over
$\sim $ 60-90~s time intervals. We recorded up to $5\sigma$  flux deviations from the mean level over
a 30-s time interval. During this time, the optical component was at maximum light:
($(JD2447677.+);$ W=~14$^{m}$.390; B=~14$^{m}$.833;
V=~14$^{m}$.338; R=~14$^{m}$.111; $\varphi$=0.761) and
($(JD2448810.+);$ W=~14$^{m}$.342; B=~14$^{m}$.821; V=~14$^{m}$.323; R=~14$^{m}$.068;
$\varphi$=0.861) in the 1989 and 1992 seasons, respectively. The observer sees both components of the CBS in the
sky plane. With such
a small temporal resolution (30-40 second exposures in the W and B  filters) we see rapid chaotic
variability with an amplitude of 0$^{m}$.08 $\div$ 0$^{m}$.10. Such a rapid variability provides conclusive
evidence for the small size of the emitting region, which should be within the  (0.5 $\div$
1.0)$\times$$10^{12}$~cm (whatever the model we adopt for the CBS).

~ A comparative analysis of the orbital-phase dependence of the W-, B-, V-, and R-band magnitudes
shows that the amplitude of rapid variability depends strongly on wavelength: it increases with decreasing
wavelength and is maximal in the W band, where it sometimes amounts to 0$^{m}$.85. Rapid variability may
result in  rather large scatter of individual observational data points on the  (W-B) - B and (B-V) - V
color-magnitude curves.

~ The system's rapid irregular variability can be naturally linked to the substantial variation of the
x-ray flux, which heats the outer regions of the accretion disk and the  ``hot spot''. The total amplitude
of irregular light variations of the  system ranged from about  $\Delta$ 1$^m$.15 in the W filter
to $\Delta$ 0$^m$.85 in the R filter. At maximum light the system exhibited chaotic variations with the
amplitude ranging from 0$^{m}$.085 to 0$^{m}$.190 in the W filter (photometric measurements are
accurate to within $\sim $ 0$^{m}$.005 - 0$^{m}$.007). At minimum light the amplitude of rapid variability
was of about $\sim $ 0$^{m}$.03 - 0$^{m}$.13 in observations with a temporal resolution of  60-90~s. Observational
errors were of about  $\sim $ 0$^{m}$.005 and $\sim $ 0$^{m}$.007 at maximum and minimum light, respectively.
The large scatter of individual observational data points must be due to rapid flux fluctuations during
a single WBVR exposure. The characteristic duration of a single WBVR exposure was of about 8-10~minutes.

\section*{Comparison of our Optical Observations to the EXOSAT and Ginga X-Ray Data}

~ We noted that may be in one of the three apparent states when  observed at optical wavelengths:
quiescent, outburst, or intermediate between the former two. The system may spend up to several days in the
intermediate stage~\cite{Sazonov1988}; ~\cite{Sazonov2006}.

~ Wijnands~\cite{Wijnands1997} reports x-ray observations (June 1987; June and October 1988; November
1990; May--June 1991) for the periods when the system was in high and intermediate states. Our optical
observations, some of which were obtained during the above periods, also exhibit two --- high and
intermediate --- states. Kuulkers~\cite{Kuulkers1996a} reports the results of a
comparative analysis of x-ray observations for the same observing periods. Optical data weakly correlate with
x-ray data at about $10\%$ level.

\section*{Amplitude of Regular Orbital Oscillations}

~ It is evident from the plots that, on the average, the object becomes redder when its flux decreases
(the (B-V) color index increases with increasing V). The object exhibits predictable behavior when its
V-band magnitude lies in the 14$^{m}$.4---15$^{m}$.0 interval, and its color indices begin to decrease
with further increase of V.

~ The plots for each observing season show appreciable variations of the $\varphi$-phased W, B, V, and R
light curves and in the (W-B) -- (B-V) and (V-R) --- (B-V) color-color curves, which appear to be due
to heating of the part of the surface of the F-type star that faces the x-ray source (the ``reflection
effect''), which show up as waves in the color-index curves with minima in the vicinity of the zero
orbital phase of the CBS.

\section*{Analysis of the Color Indices}

~ The variations of the W-B color index have an amplitude of about 0$^{m}$.42. Maximum orbital light
variations show up conspicuously with the amplitudes of  $\Delta$ 0$^{m}$.30, $\Delta$ 0$^{m}$.278,
$\Delta$ 0$^{m}$.265, and $\Delta$ 0$^{m}$.26 in the W, B, V, and R-band filters, respectively, and with
sharp light minima at two conjunctions.

The minimum near the phase $\varphi$= 0$^{p}$.50 is sharper, because, according to spectroscopic data,
the F-type star passes the periastron of its orbit near the upper conjunction, and spends little
time at these phases$\prime $ due to the elliptical shape of its orbit.

\section*{Evolution and Position of the CBS V 1341 Cyg = Cyg X-2 on the Two-Color Diagram}

~ We now summarize the crucial points of this section (and of the paper as a whole!):

(1) Note that an important conclusion made by Kuznetsov~\cite{Kuznezov2002} requires a substantial correction
to be applied to the previous distance estimates for Cyg X-2, which are based on optical observations  (and employed
by Goranskii~\cite{Goranskii1988} : the estimate of Cowley~\cite{Cowley1979} differs by $\sim ${30-60} percent
(d=8.7$^{-1.8}_{+2.2})$.

(2) In the two-color evolutionary diagram the optical component of the CBS V1341 Cyg is located in the
region of the turnoff of the MS of globular cluster stars (for the adopted E$_{B-V}$), implying that the object
should be a dwarf or a star near the point of turnoff toward red giants.

~ Hence the points (1) and (2) just mentioned and our observations and their interpretation are inconsistent
with the assumption that the primary component is a  «blue straggler».

\section*{Contribution of the Accretion Disk and Accretion Formations to the Total Flux of the System}

~ Our qualitative assessment of the contribution of the flux of the accretion disk takes into account the quantitative
data about the position of the object on the (W-B)--(B-V) two-color diagram during different states of the
system. Near the primary minimum Min I (JD2447414, JD2447444) the system exhibits chaotic variability
with an amplitude ranging from 0$^{m}$.085 to 0$^{m}$.190. The accuracy of W-band photometry is of about
$\sim $ 0$^{m}$.01.

~ We already pointed out above that UV flux outbursts are observed near orbital phases $\varphi$=0$^p$.3 and
$\varphi$=0$^p$.8 even when the system is in the quiescent state. These high-temperature flux outbursts
(they are less conspicuous in the  V and R bands!) must apparently be interpreted in terms of the
\textbf{transition layer model (TLM)}.

~ The contribution of the accretion disk to the total flux of the system varies from about $4-5 \%$ (quiescent
state) to about $50 \%$ (active state of the system).

\end{document}